\documentclass[aps,prd,preprint,superscriptaddress,tightenlines,floatfix,showpacs]{revtex4-1}
\usepackage{graphics}
\usepackage{graphicx}
\usepackage{amsmath,amssymb}
\usepackage{tabularx}
\usepackage[english,activeacute] {babel}

\begin{document}

\title{Black holes in asymptotic safety with higher derivatives: accretion and stability analysis}

\author{Fabi\'an H. Zuluaga}\email{fhzuluagag@unal.edu.co}

\author{Luis A. S\'anchez}\email{lasanche@unal.edu.co}

\affiliation{Escuela de F\'\i sica, Universidad Nacional de Colombia,
A.A. 3840, Medell\'\i n, Colombia}

\begin{abstract}
We review the issue of steady spherically symmetric accretion onto a renormalization group improved Schwarzschild space-time which is solution to an asymptotically safe theory (AS) containing high-derivative terms. We use a Hamiltonian dynamical system approach for the analysis of the accretion of four types of isothermal test fluids: ultra-stiff fluid, ultra-relativistic fluid, radiation fluid, and sub-relativistic fluid. An important outcome of our study is that, contrary to what claimed in a recent work, there exist physical solutions for the accretion of an ultra-relativistic fluid in AS which include subsonic, supersonic and transonic regimes. We also study quantum corrections to the known stability of the accretion in general relativity (GR). To this end we use a perturbative procedure based on the continuity equation with the mass accretion rate being the perturbed quantity. Two classes of perturbations are studied: standing waves and travelling waves. We find that quantum gravity effects either enhance or diminish the stability of the accretion depending on the type of test fluid and depending on the radial distance to the central object.\\

\noindent
{\bf keywords}: asymptotic safety, black holes, accretion, stability.
\end{abstract}


\maketitle
\section{\label{sec:intr}Introduction}
The event horizon of a black hole is defined in general relativity as the frontier of a region of space-time where gravity is so strong that nothing, not even light, that enters that region can ever escape. The characteristic features of black holes, including event horizons, accretion and outflow processes, singularities, Hawking radiation, etc., and the possibility of observation of their immediate environment with angular resolution comparable to the event horizon using facilities such as the Event Horizon Telescope \cite{1} and the GRAVITY instrument \cite{2}, make black holes useful laboratories to test general relativity in the strong field regime and, hopefully, to test ideas for quantum gravity. 

Accretion process onto compact objects, on the other hand, is relevant since it is believed to provide the energy of Quasars, accreting supermassive black holes, X-ray binaries and Gamma-ray bursts. The simplest situation, which has reached a paradigmatic status in studies on accretion astrophysics, is spherically symmetric accretion onto a static compact object of inviscid gas which is at rest at infinity. In the framework of Newtonian gravity the problem was analyzed in a seminal paper by Bondi \cite{3-Bondi}. The relativistic generalization of the Bondi accretion problem was done in \cite{4-Michel}, which was followed by a large number of studies analyzing spherical accretion onto a wide variety of astrophysical objects including black holes \cite{5-various}. 

The problem of the stability of spherically symmetric accretion under linearized perturbations has extensively been studied in the literature both in the context of Newtonian gravity \cite{6-PETT,7-TD,8-Garlick,9-Gaite} and in the classical general relativistic realm \cite{10-Moncrief,11-Malec,12-DS,13-MRD,14-Ray,15-RR,16-MM,17-Naskar}. In both of these regimes no evidence has been found of the development of instabilities. In \cite{6-PETT} it has been shown that in the Newtonian regime and neglecting viscosity, the perturbation has a constant amplitude for the conserved flow so that stability is ensured. In general relativity the background solution is also stable with the amplitude of the perturbation being damped in time, that is, general relativistic effects enhance the stability \cite{10-Moncrief}, \cite{17-Naskar}. A mechanism that favors stability in this case is the coupling of the infalling flow with the geometry of space-time, which acts in the manner of a dissipative effect.

In this work we consider quantum gravity corrections to the accretion, and to its stability, onto an improved Schwarzschild black hole in the framework of the asymptotic safety scenario for quantum gravity. In this scenario it is conjectured that gravity constitutes a fundamental theory at the non-perturbative level \cite{18-Weinberg}. A basic ingredient in this construction is a non-Gaussian fixed point (NGFP) of the gravitational renormalization group (RG) flow which controls the behavior of the theory at trans-Planckian energies, where the physical degrees of freedom interact predominantly antiscreening, and that renders physical quantities safe from unphysical divergences at all scales \cite{19-Reuter}. This means that the gravitational field itself is the principal carrier of the relevant classical and quantum degrees of freedom, with the renormalization flow relating its infrared and ultraviolet physics. On condition that the set of RG trajectories approaching the fixed point in the UV are parameterized by a finite number of relevant (that is, physically motivated) running couplings, asymptotic safety is as predictive as the standard perturbatively renormalizable quantum field theory. Notwithstanding asymptotic safety defines a consistent and predictive quantum theory for gravity within the framework of quantum field theory, it remains a prediction in the sense that a rigorous existence proof for the NGFP is still lacking. There exist, however, substantial evidence for the existence of such a fixed point suitable for the Weinberg's asymptotic safety scenario \cite{20-various2}. 

Quantum corrections to the structure of black hole spacetimes within the asymptotic safety program was initiated in \cite{21-BoRe} by constructing the so-called ``RG-improved'' solution to the Schwarzschild metric. Following up on these initial work, the RG-improved Kerr metric has been discussed in \cite{22-ReTu}, and the improved Schwarzschild-(A)dS and Reissner-Nordstr\"{o}m-(A)dS metrics have been studied in \cite{23-Ko-Sa} and \cite{24-Go-Ko}, respectively. The inclusion of a cosmological constant is relevant because it has been shown in \cite{23-Ko-Sa} that is this constant which determines the short-distance structure of the RG-improved black holes. Further aspects of these quantum compact objects including thermodynamical properties, evaporation processes, mini-black hole production, etc, have been analyzed in \cite{25-various3}, while quantum corrections to the accretion process onto a Schwarzschild black hole have been discussed in \cite{26-Yang}, \cite{27-FAYJ}. 

In this paper we review the results by the authors in \cite{27-FAYJ} where the study of the accretion onto an improved asymptotically safe Schwarzschild metric with high-derivative terms was done, and we also discuss the stability of the accretion process and contrast with the result coming from GR. The interest in the improved Schwarzschild space-time with higher-derivatives in the infrared limit comes from the fact that, as shown in \cite{28-Cai-Easson}, describes a black hole that never completely evaporate and, consequently, makes an excellent dark matter candidate. Since our interest will be to analyze the most general aspects of the problem, we will neglect viscous effects, heat transport, fluid self-gravity, and effects associated to the back-reaction of the fluid on the geometry. As done in \cite{27-FAYJ}, we describe the accretion process of isothermal fluids by using the Hamiltonian dynamical formalism and we show that, contrary to what claimed in that work, there exist physical solutions for the accretion of an ultra-relavitivistic fluid in AS. Regarding the analysis of the stability, instead of adopting the approach of perturbing a scalar potential whose gradient is prescribed to be the velocity of the ideal fluid \cite{10-Moncrief}, we use a perturbation scheme based on the continuity equation \cite{6-PETT}, \cite{7-TD}, \cite{17-Naskar}. The stationary solution of the continuity equation gives a first integral, which, within a constant geometric factor, is actually the matter accretion rate. Our perturbative procedure entails the perturbation of this constant of motion. This quantity, being a flux of mass can be observed and, in principle, can be measured using the present day and the projected observational instruments \cite{1}, \cite{2}.

This paper is organized as follows. In the next section we present the RG-improved Schwarzschild metric. In Sec.~\ref{sec:sec3} we review the mathematics of the accretion process and present the Hamiltonian dynamical system formalism. In Sec.~\ref{sec:sec4} we apply the formalism to the spherically symmetric accretion of isothermal fluids and analyze the quantum gravity effects by comparing with the GR description. In Sec.~\ref{sec:sec5} we discuss the implementation of the perturbation scheme based on the continuity equation, and analyze the quantum gravity corrections to the stability of the accretion by studying the behavior of standing wave and travelling wave perturbations to the mass accretion rate. Finally, in the last section, we present our conclusions.
%
\section{\label{sec:sec2}Renormalization group improved spherically symmetric space-time with higher derivatives}
We start recalling that the classical Schwarzschild line element for a black hole with mass $M$, in units $c =G_{0}= 1$, is written as
\begin{equation}\label{eq1}
ds^2=-f_0(r)dt^2+f_0(r)^{-1}dr^2+r^2\Big(d\theta^2+\sin^2\theta d\varphi^2\Big),
\end{equation}
with the metric coefficient $g^{tt}$ given by: $f_0(r) = 1-2M/r$. The event horizon or Schwarzschild radius comes from solving $f_0(r)=0$, which gives $r_{hS}=2M$.

\begin{figure}[!ht]
	\centering
	\minipage{0.45\textwidth}
	\includegraphics[width=6.2cm,height=5.0cm]{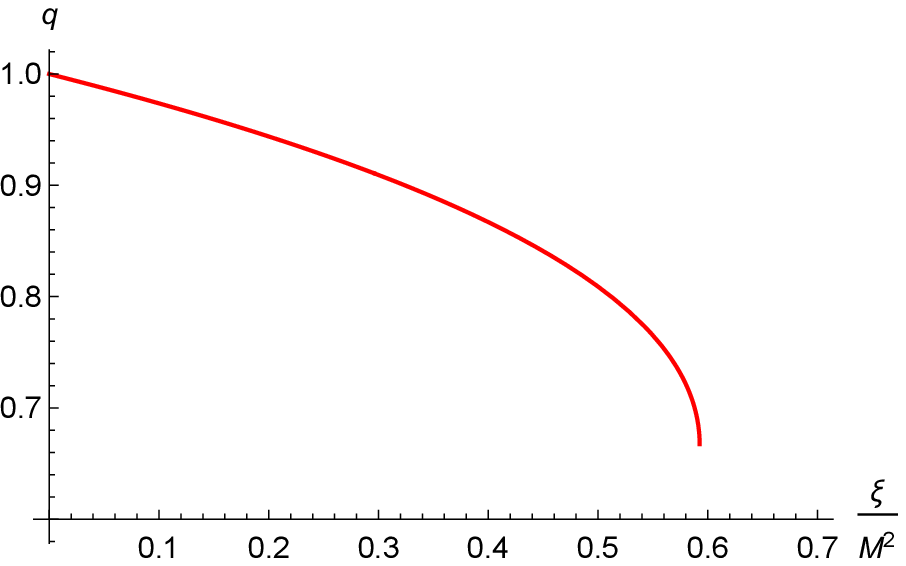}
	\label{f1_1}
	\endminipage\hfill
	\minipage{0.45\textwidth}
	\includegraphics[width=6.4cm,height=5.0cm]{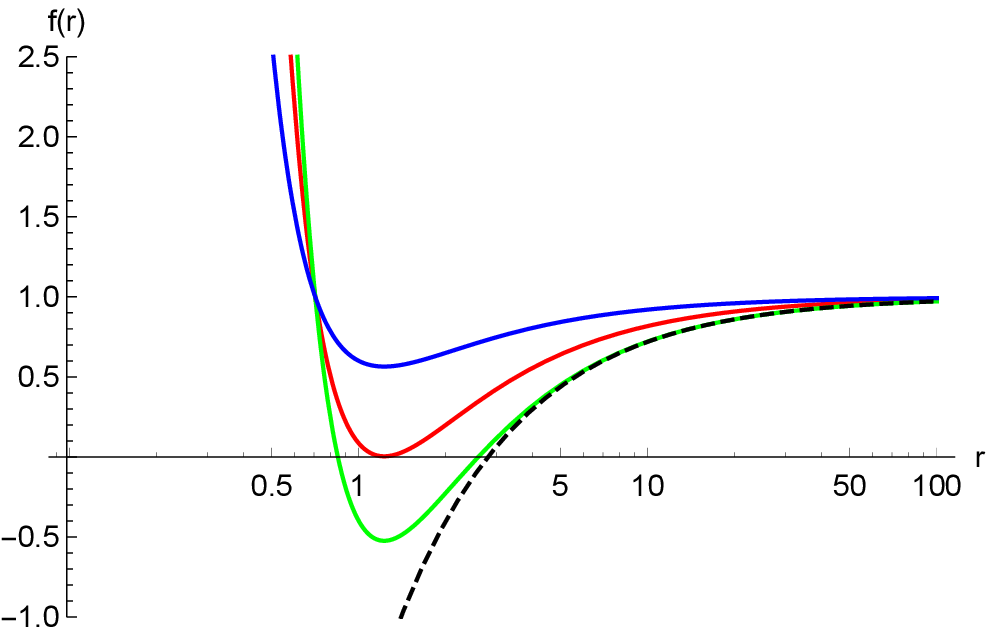}
	\label{f1_2}
	\endminipage\hfill\\
	\caption{Left panel: Dependence of $q\equiv r_{hIR}/r_{hS}$ on $\tilde{\xi}=\xi/M^2$. Rigt panel: plots of the improved metric coefficient $f(r)$ for $M<M_{c}$ (blue), $M=M_{c}$ (red), and $M>M_{c}$ (green), with $M_{c}=0.918$ corresponding to $\xi = 0.5$. The dashed line correspond to the classical $f_0(r)$ for $M=M_{c}$.} \label{Fig1}
\end{figure}

On the other hand, in the high-derivative gravity theory, which includes the Ricci scalar square, the Ricci tensor square, the Kretschmann scalar and running gravitational couplings in the effective action \cite{28-Cai-Easson}, the quantum gravity corrections to the metric are accounted for by promoting the classical gravitational Newton constant through a running coupling that evolve under the equations of the gravitational RG-flow. Thus, in the low energy limit the RG-improvement to the metric in Eq.~(\ref{eq1}) is obtained by doing
\begin{equation}\label{eq2}
f_0(r) \rightarrow f(r) = 1-\frac{2M}{r}\left(1-\frac{\xi}{r^2}\right),  
\end{equation}
where $\xi$ is a parameter with dimensions of length squared associated to the scale identification between the momentum scale $p$ and the radial coordinate $r$ which, in the IR regime, takes the form $p \sim 1/r$. Note that the quantum effects on the space-time geometry are all encoded in $\xi$, such that for $\xi=0$ the classical metric coefficient $f_0(r)$ is recovered. The condition for the radius of the new event horizon: $f(r)=0$, takes the form of a generic cubic equation
\begin{equation}\label{eq3}
r^3-2 M r^2+2 M^3 \tilde{\xi} = 0,  
\end{equation}
where the dimensionless parameter $\tilde{\xi}=\xi/M^2$ has been introduced for convenience. The only real solution $r_{hIR}$ to Eq.(\ref{eq3}) is
\begin{equation}\label{eq4}
\frac{r_{hIR}}{r_{hS}}=\frac{1}{6} \left(2+\frac{4}{\sqrt[3]{8-27 \tilde{\xi} +3 \sqrt{3} \sqrt{\tilde{\xi} (27 \tilde{\xi} -16)}}}+\sqrt[3]{8-27 \tilde{\xi} +3 \sqrt{3} \sqrt{\tilde{\xi} (27 \tilde{\xi} -16)}}\right).  
\end{equation}
The left panel in Fig.~\ref{Fig1} shows the quotient $q\equiv r_{hIR}/r_{hS}$ as a function of $\tilde{\xi}$. For $\tilde{\xi}=0$ we have $r_{hIR}=r_{hS}$ as expected, while for $\tilde{\xi}$ greater than the critical value $\tilde{\xi}_{c}=16/27$, there is no horizon at all and a naked singularity arises. This implies that each value of $\tilde{\xi}$ in the range $0 \leq \tilde{\xi} \leq \tilde{\xi}_{c}$ picks out a critical mass $M_{c}=\sqrt{27\xi/16}$ in such a way that for $M>M_{c}$ there are two horizons: one inner Cauchy horizon and one outer event horizon, for $M<M_{c}$ a naked singularity develops, while when $M=M_{c}$ the two horizons merge. The right panel in Fig.~\ref{Fig1} illustrates this situation for $\xi=0.5$ that is, for $M_c = 0.918$. Since our aim is to discuss quantum gravity effects on the accretion onto a Schwarzschild black hole, naked space-time singularities will be ignored.

For future reference we note that for $M>M_{\mathrm{c}}$, the outer horizon of the improved solution, upon expanding to the leading order in $\xi$, acquires the form: $r_{hIR}=r_{hS}-\xi /(2M)$. This result exhibits the typical shifting of the horizon of RG-improved black hole solutions towards smaller values with respect to its classical counterpart.

\section{\label{sec:sec3}Spherically symmetric accretion as a dynamical system}
Here we describe the accretion flow by adopting the Hamiltonian procedure. The formulation of the problem of accretion onto compact objects as a Hamiltonian dynamical system was proposed for the first time in \cite{29-CHA-SA}.
 
We first recall that the phenomenon of accretion is based on two conservation laws: the continuity equation which expresses the conservation of the particle number, and the energy-momentum conservation. These requirements are given, respectively, by the covariant derivatives
\begin{equation}\label{eq5}
\nabla_{\mu} \left(\eta u^{\mu} \right)=0  \qquad \mbox{and} \qquad \nabla_{\mu} T^{\mu \nu} = 0,
\end{equation}
with $\eta$ being the particle number density of the fluid, and $u^{\mu}$ the four-velocity of the fluid normalized as $u^{\mu}u_{\mu}=-1$.

Neglecting effects related to viscosity or heat transport, and assuming that the fluid's energy density is sufficiently small so that its self-gravity can also be ignored, the accreting matter can be approximated as a perfect fluid described by the energy-momentum tensor
\begin{equation}\label{eq6}
T_{\mu \nu} = \left( \epsilon + p \right) u_{\mu} u_{\nu} + pg_{\mu \nu},
\end{equation}
where $\epsilon$ and $p$ are the proper energy density and the proper pressure of the fluid, respectively. 

For purely radial flow in the equatorial plane the normalization condition $u^{\mu}u_{\mu}=-1$ produces
\begin{equation}\label{eq7}
u_t = -f u^t = \sqrt{f+(u^r)^2}
\end{equation}
where $u^r$ denotes the radial flow velocity and $f=f(r)$ is the metric coefficient in the spherically symmetric line element. 

We assume as infalling matter a test fluid which is both isothermal and isentropic, that is a test fluid with barotropic equation of state $h=h(\eta)$ where $h$ is the specific enthalpy
\begin{equation}\label{eq8}
h=\frac{\epsilon + p}{\eta}.
\end{equation}
This fluid is described by the thermodynamical equations
\begin{eqnarray}\label{eq9}
dp = \eta dh, \qquad d\epsilon = hd\eta,
\end{eqnarray}
so that the square of the local speed of sound is written as
\begin{equation}\label{eq10}
a^{2} = \frac{\partial p}{\partial \epsilon}\Bigg\vert_{ds=0}=\frac{dp}{d\epsilon}=\frac{\eta}{h}\frac{dh}{d\eta}=\frac{d \ln h}{d \ln \eta}.
\end{equation}
Then, as it is well known, for spherically symmetric stationary accretion the conservation laws in Eq.~(\ref{eq3}) yield 
\begin{equation}\label{eq11}
\eta u^r r^2 = C_1,
\end{equation}
and
\begin{equation}\label{eq12}
h u_t = C_2,
\end{equation}
where $C_1$ and $C_2$ are arbitrary integration constants.

In order to describe the accretion process as a two-dimensional Hamiltonian dynamical system, it is useful to define the three-velocity $v$ of a fluid element as measured by a locally static observer as $v=dl/d\tau$, where $dl=dr/\sqrt{f}$ and $d\tau = \sqrt{f} dt$ are the proper radial distance and the proper time, respectively. It is important to remark that $v$ is defined outside the horizon, that is for an observer for which the time coordinate is timelike \cite{30-AHMED}. Using $u^r=dr/d\tau$, $u^t=dt/d\tau$ and Eq.~(\ref{eq7}) we get
\begin{equation}\label{eq13}
v^2 = \frac{(u^r)^2}{f+(u^r)^2}=\frac{(u^r)^2}{(u_t)^2},
\end{equation}
from which
\begin{equation}\label{eq14}
(u^r)^2 = \frac{fv^2}{1-v^2}, \quad \mathrm{and} \quad (u_t)^2=\frac{f}{1-v^2}.
\end{equation}
As the Hamiltonian for the flow fluid we can choose either of the two integrals of motion $C_1, C_2$, or even any combination of them. We choose $C_2^2$ as our Hamiltonian $H$ and fix the dynamical variables of the system to be $r$ and $v$. Using Eq.~(\ref{eq14}) we then have
\begin{equation}\label{eq15}
H(r,v)= \frac{h^2(r,v)f(r)}{1-v^2},
\end{equation}
and the dynamical system is defined by
\begin{equation}\label{eq16}
\dot{r}=\frac{\partial H}{\partial v}, \qquad   \dot{v}=-\frac{\partial H}{\partial r},
\end{equation}
where the dot denotes time derivative. After using Eq.~(\ref{eq10}) these two equations become
\begin{equation}\label{eq17}
\dot{r}=\frac{2 f h^2}{v(1-v^2)^2}(v^2-a^2),,
\end{equation}
\begin{equation}\label{eq18}
\dot{v}=-\frac{h^2}{r(1-v^2)}\left[(1-a^2)r \frac{df}{dr}-4 f a^2\right]
\end{equation}
The critical points of the dynamical system, which coincides with the sonic points of the fluid flow, are the points $(r_c,v_c)$ where both $\dot{r}$ and $\dot{v}$ are zero. Assuming that $h\neq 0$ for all values of $r>r_{hIR}$ where also $f\neq 0$, it follows that at the critical points
\begin{equation}\label{eq19}
v_c^2=a_c^2, \qquad r_c(1-a_c^2)f_{c,r_c} = 4f_c a_c^2,
\end{equation}
where $f_c=f(r)\vert_{r_c}$ and $f_{c,r_c}=(df/dr)\vert_{r_c}$.

\section{\label{sec:sec4}Accretion of isothermal fluids}
In this section, following \cite{30-AHMED}, we will obtain an adequate expression for $h(r,v)$ in Eq.~(\ref{eq15}) when the accreted fluid is isothermal. This in turn will provide a neat expression for the Hamiltonian $H(r,v)$.

The barotropic equation of state for an isentropic fluid can be expressed as $\epsilon=\epsilon(\eta)$. Moreover, for an isothermal fluid the equation of state reads $p=k\epsilon$ where the constant $k$ is the parameter state: $0<k<1$. Note that the definition of the speed of sound in Eq.~(\ref{eq10}) leads to $a^2=k$ and so the speed of sound remains constant through the accretion process. Now, from Eqs.~(\ref{eq9}) we have
\begin{equation}\label{eq20}
h=\frac{d\epsilon}{d\eta},
\end{equation}
and
\begin{equation}\label{eq21}
\frac{dp}{d\eta}=\eta \frac{dh}{d\eta}=\frac{d^2\epsilon}{d\eta^2},
\end{equation}
which, upon integration, produces
\begin{equation}\label{eq22}
\eta \frac{d\epsilon}{d\eta}-\epsilon(\eta)=k\epsilon(\eta),
\end{equation}
where we have used $p(\eta)=k\epsilon(\eta)$. Integration of this last equation yields
\begin{equation}\label{eq23}
\epsilon(\eta)=C \eta^{k+1}=\frac{\epsilon_c}{\eta_c^{k+1}}\eta^{k+1},
\end{equation}
where the integration constant $C$ has been chosen so that Eqs.~(\ref{eq8}) and (\ref{eq20}) give rise to the same expression for $h$
\begin{equation}\label{eq24}
h(r,\eta)=\frac{(k+1)\epsilon_c}{\eta_c}\left(\frac{\eta}{\eta_c} \right)^k,
\end{equation}
An expression for $\eta/\eta_c$ can be obtained from the constant of integration $C_1$ in  Eq.~(\ref{eq11}). In fact, using Eqs.~(\ref{eq14}) and (\ref{eq19}) we can write
\begin{equation}\label{eq25}
C_1^2=\frac{r^4 \eta^2 f v^2}{1-v^2}=\frac{r_c^4 \eta_c^2 f_c v_c^2}{1-v_c^2}=\frac{r_c^5 \eta_c^2 f_{c,r_c}}{4},
\end{equation}
so that
\begin{equation}\label{eq26}
\left(\frac{\eta}{\eta_c} \right)^k=\left(\frac{r_c^5 f_{c,r_c}}{4}\frac{1-v^2}{r^4 f v^2} \right)^{k/2}.
\end{equation}
Replacing into Eq.~(\ref{eq24}) we obtain
\begin{equation}\label{eq27}
h^2= K \left(\frac{1-v^2}{r^4 f v^2}\right)^k,
\end{equation}
with the constant $K$ given by $K=(r_c^5 f_{c,r_c}/4)^k[(k+1)\epsilon_c/\eta_c]^2$.

Redefining the Hamiltonian as $\mathcal{H}=H/K$ and using Eq.~(\ref{eq26}), $\mathcal{H}$ acquires finally the form
\begin{equation}\label{eq28}
\mathcal{H}(r,v)=\frac{[f(r)]^{1-k}}{r^{4k}v^{2k}(1-v^{2})^{1-k}}.
\end{equation}
It must be stressed that, due to the definition of the three-velocity $v$, this expression for $\mathcal{H}$ is valid for an observer outside the horizon. This applies both for the classical Schwarzschild metric and for the improved AS metric. 

We now focus on four types of isothermal accretion: ultra-stiff fluid ($k=1$), ultra-relativistic fluid ($k=1/2$), radiation fluid ($k=1/3$), and sub-relativistic fluid ($k=1/4$). For the purpose of comparison with the analysis in \cite{27-FAYJ}, we will take $\xi =0.5$ and $M=1>M_{c}=0.918$ in the expressions for the quantum $\mathcal{H}^{(AS)}$ and the classical $\mathcal{H}^{(GR)}$ Hamiltonians. Notwithstanding this amounts to consider a black hole solution with two horizons, the quantum effects on the accretion we will describe below are present for all the values of the parameter $\xi$ in the range $0\leq \xi \leq (16/27)M^2$, with the orbits in the contour plot of the quantum Hamiltonian going continuously towards the orbits of its classical counterpart in the limit $\xi \rightarrow 0$ as we will show ahead.

\subsection{\label{sec:sec4.1}Solutions for $k=1$}
The equation of state for ultra-stiff fluid reads $p=\epsilon$, which implies $a^2=1=v_c^2$. Eq.~(\ref{eq19}) immediately gives $f_c=0$ and the Hamiltonian in AS coincides with the one in GR
\begin{equation}\label{eq29}
\mathcal{H}^{(AS)}=\mathcal{H}^{(GR)}=\frac{1}{r^{4}v^{2}},
\end{equation}
with the only difference being in the location of the horizon that, in both cases, coincides with the location of the critical point: $r_c^{(AS)}=r_{hIR}$ and $r_c^{(GR)}=r_{hS}$.

\begin{figure}[!ht]
\centering
\includegraphics[scale=0.6]{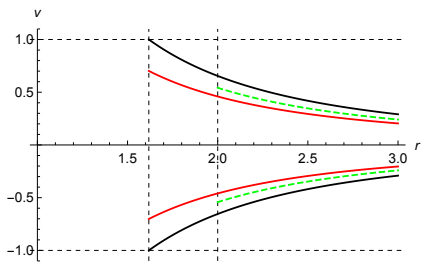}
\caption{\label{Fig2} 
Contour lines of $\mathcal{H}^{(AS)}=\mathcal{H}_{c}^{(AS)}$ (black) and $\mathcal{H}^{(AS)}=\mathcal{H}_{c}^{(AS)}\pm 0.06$ (red) for accretion of ultra-stiff fluid ($k=1$). The green dashed curve is the contour line of $\mathcal{H}^{(GR)}=\mathcal{H}_{c}^{(GR)}\pm 0.06$. $\mathcal{H}_{c}^{(AS,GR)}$ stands for $\mathcal{H}^{(AS,GR)}(r)\vert_{r_c}$. For all the orbits: $M=1$, $\xi=0.5$. The vertical dashed lines locate the horizons $r_{hIR}$ (left) and $r_{hS}$ (right). The horizontal dashed lines locate the speed of sound $a=\sqrt{k}$.
 }
\end{figure}

Contour plots of $\mathcal{H}^{(AS)}$ (continuous lines) and of $\mathcal{H}^{(GR)}$ (dashed line) are shown in Fig.~\ref{Fig2} where we have only included selected orbits associated to physical flow ($\vert v \vert <1$). Here we retrieve the findings in \cite{27-FAYJ} concerning the contour lines for $\mathcal{H}^{(AS)}$. Note that both for infalling matter ($-1<v<0$) and for fluid outflow ($0<v<1$), and for fixed values of the radial coordinate $r$ with $r>r_{hS}$, the flow velocity is always greater in GR than in AS. 

We note that we have restricted the comparison between classical and quantum accretion and outflow to the region $r>r_{hS}$ in the plots because this is the region for which the comparison makes sense.
%

\subsection{\label{sec:sec4.2}Solutions for $k=1/2$}
In the case of an ultra-relativistic fluid we have $a^2=1/2=v_c^2$, and the Hamiltonians reads
\begin{eqnarray}\nonumber
\mathcal{H}^{(AS)}&=&\frac{\sqrt{1-\frac{2M}{r}+\frac{2M\xi}{r^3}}}{r^{2}v\sqrt{1-v^{2}}}, \\ \label{eq30} 
\mathcal{H}^{(GR)}&=&\frac{\sqrt{1-\frac{2M}{r}}}{r^{2}v\sqrt{1-v^{2}}}.
\end{eqnarray}
For the critical radii we get
\begin{equation}\label{eq31}
r_c^{(AS)}\simeq \frac{5}{2} M - \frac{14}{25}\frac{\xi}{M}, \qquad r_c^{(GR)}= \frac{5}{2}M,
\end{equation}
where $r_c^{(AS)}$ is given at leading order in $\xi$.

\begin{figure}[!ht]
\centering
\includegraphics[scale=0.6]{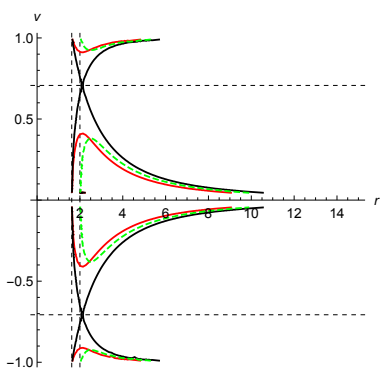}
\caption{\label{Fig3} 
Contour lines of $\mathcal{H}^{(AS)}=\mathcal{H}_{c}^{(AS)}$ (black) and $\mathcal{H}^{(AS)}=\mathcal{H}_{c}^{(AS)}\pm 0.06$ (red) for accretion of ultra-relativistic fluid: $k=1/2$. The green dashed curve is the contour line of $\mathcal{H}^{(GR)}=\mathcal{H}_{c}^{(GR)}\pm 0.06$. For all the orbits: $M=1$, $\xi=0.5$. The vertical dashed lines locate the horizons $r_{hIR}$ (left) and $r_{hS}$ (right). The horizontal dashed lines locate the speed of sound $a=\sqrt{k}$.
 }
\end{figure}

In Fig.~\ref{Fig3} the contour plots of $\mathcal{H}^{(AS)}$ (continuous lines) and of $\mathcal{H}^{(GR)}$ (dashed line) are shown. By readability of the figure we have not depicted the contour line $\mathcal{H}^{(GR)}=\mathcal{H}_{c}^{(GR)}$. The important fact here is the existence of orbits associated to physical flow ($\vert v \vert <1$) in AS that include subsonic regime ($-v_c<v<v_c$), supersonic regime ($-v<-v_c$ and $v>v_c$) and transonic regime associated to the solution passing through the sonic point. This is the opposite to the findings by the authors in \cite{27-FAYJ} where no physical significance was established for accretion of an ultra-relativistic fluid in AS. Two quantum gravity effects can be identified: a shifting of the orbits towards the black hole and an increase of the maximum flow velocity, which occurs at $r=r_c$, as compared with the case in GR.

\subsection{\label{sec:sec4.3}Solutions for $k=1/3$}
\begin{figure}[!ht]
\centering
\includegraphics[scale=0.45]{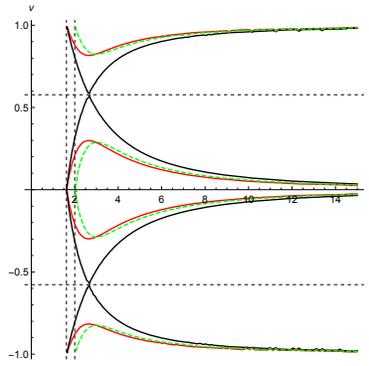}
\caption{\label{Fig4} 
Contour lines of $\mathcal{H}^{(AS)}=\mathcal{H}_{c}^{(AS)}$ (black) and $\mathcal{H}^{(AS)}=\mathcal{H}_{c}^{(AS)}\pm 0.06$ (red) for accretion of radiation: $k=1/3$. The green dashed curve is the contour line of $\mathcal{H}^{(GR)}=\mathcal{H}_{c}^{(GR)}\pm 0.06$. For all the orbits: $M=1$, $\xi=0.5$. The vertical dashed lines locate the horizons $r_{hIR}$ (left) and $r_{hS}$ (right). The horizontal dashed lines locate the speed of sound $a=\sqrt{k}$.
 }
\end{figure}

For radiation fluid the critical radii are given by
\begin{equation}\label{eq32}
r_c^{(AS)} \simeq 3 M-\frac{5}{9}\frac{\xi}{M}, \qquad r_c^{(GR)} = 3 M,
\end{equation}
while the AS and RG Hamiltonians takes respectively the form
\begin{eqnarray}\nonumber
\mathcal{H}^{(AS)}&=&\frac{\Big(1-\frac{2M}{r}+\frac{2M\xi}{r^3}\Big)^{2/3}}{r^{4/3}v^{2/3}(1-v^{2})^{2/3}}, \\ \label{eq33} 
\mathcal{H}^{(GR)}&=&\frac{\Big(1-\frac{2M}{r}\Big)^{2/3}}{r^{4/3}v^{2/3}(1-v^{2})^{2/3}}
\end{eqnarray}
The orbits of $\mathcal{H}^{(AS)}$ (continuous lines) and of $\mathcal{H}^{(GR)}$ (dashed line) are shown in Fig.~\ref{Fig4}. Here we also recover the results in \cite{27-FAYJ} concerning the contour lines for $\mathcal{H}^{(AS)}$. Notice that the overall behavior of the accretion and the outflow is the same as in the case $k=1/2$, which implies the same effects coming from quantum gravity discussed above. 
%
\subsection{\label{sec:sec4.4}Solutions for $k=1/4$}
\begin{figure}[!ht]
\centering
\includegraphics[scale=0.6]{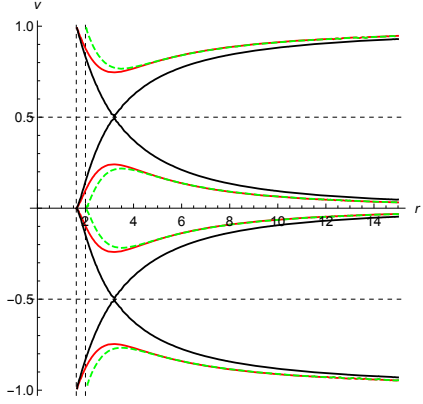}
\caption{\label{Fig5} 
Contour lines of $\mathcal{H}^{(AS)}=\mathcal{H}_{c}^{(AS)}$ (black) and $\mathcal{H}^{(AS)}=\mathcal{H}_{c}^{(AS)}\pm 0.06$ (red) for accretion of sub-relativistic fluid: $k=1/4$. The green dashed curve is the contour line of $\mathcal{H}^{(GR)}=\mathcal{H}_{c}^{(GR)}\pm 0.06$. For all the orbits: $M=1$, $\xi=0.5$. The vertical dashed lines locate the horizons $r_{hIR}$ (left) and $r_{hS}$ (right). The horizontal dashed lines locate the speed of sound $a=\sqrt{k}$.
 }
\end{figure}

For sub-relativistic fluid we have
\begin{equation}\label{eq34}
r_c^{(AS)}\simeq \frac{7}{2}M-\frac{26}{49}\frac{\xi}{M}, \qquad r_c^{(GR)}=\frac{7}{2}M
\end{equation}
and
\begin{eqnarray}\nonumber
\mathcal{H}^{(AS)}&=&\frac{\Big(1-\frac{2M}{r}+\frac{2M\xi}{r^3}\Big)^{3/4}}{rv^{1/2}(1-v^{2})^{3/4}}, \\ \label{eq35} 
\mathcal{H}^{(GR)}&=&\frac{\Big(1-\frac{2M}{r}\Big)^{3/4}}{rv^{1/2}(1-v^{2})^{3/4}}
\end{eqnarray}
The contour plots for $\mathcal{H}^{(AS)}$ (continuous lines) and of $\mathcal{H}^{(GR)}$ (dashed line) in Fig.~\ref{Fig5} show that, once again, the overall behavior of the contour lines is the same as in the two previous cases, with the same global quantum gravity effects on the accretion and outflow processes. Notwithstanding, we observe that for large values of $r$ $(r>r_c)$ the quantum effects are milder for sub-relativistic matter than for ultra-relativistic and radiation fluids.

\begin{figure}[!ht]
\centering
\includegraphics[scale=0.8]{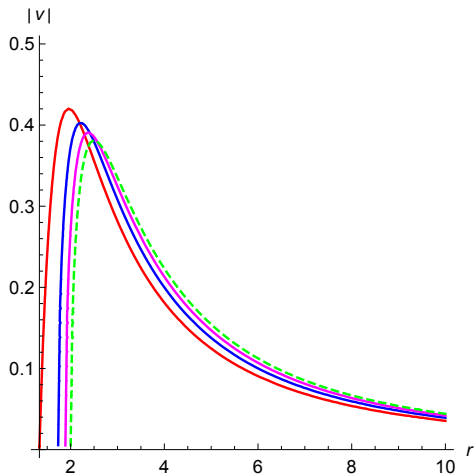}
\caption{\label{Fig6} 
Orbits of $\mathcal{H}^{(AS)}=\mathcal{H}_{c}^{(AS)}\pm 0.06$ for subsonic accretion of ultra-relativistic fluid ($k=1/2$) and for $\xi=\xi_c =16/27$ (red), $\xi=0.4$ (blue), $\xi=0.2$ (magenta). The green dashed curve is the contour line of $\mathcal{H}^{(GR)}=\mathcal{H}_{c}^{(GR)}\pm 0.06$ ($\xi=0$). For all the orbits $M=1$.
 }
\end{figure}

At this point, it is important to show more explicitly how quantum effects modify accretion, that is, the dependence of the accretion flow on the parameter $\xi$. This is displayed in Fig.~\ref{Fig6} for the case of accretion of ultra-relativistic fluid ($k=1/2$) and for subsonic flow. As can be seen, with the increase of the value of $\xi$ (from right to left in the figure), the effects of shifting of the orbit and of increasing of the maximum flow velocity, already mentioned, become more and more pronounced, whereas in the opposite limit: $\xi \rightarrow 0$, the classical orbit is retrieved.

\section{\label{sec:sec5}Stability analysis}
Let us start, following \cite{17-Naskar}, by writing the equations describing the accretion process in a more appropriate way for the stability analysis.

First, the explicit form of the continuity equation reads
\begin{equation}\label{eq36}
\partial_{t}\left(\eta u^t\right)+ r^{-2} \partial_{r} \left(\eta u^r r^2\right)=0.
\end{equation}
To solve for $\eta$ and $u^r$, we use the equation for energy-momentum conservation and the thermodynamic equation for mass-energy conservation $d(\epsilon/\eta)+P d(1/\eta)=T ds$. Along with the condition of constant entropy $ds=0$, and with the square of the speed of sound given by Eq.~(\ref{eq10}), the condition of energy-momentum conservation reads
\begin{eqnarray}\nonumber
u^t \partial_{t} u^t &+& u^r \partial_{r} u^t + f\left(\partial_{r}f\right)u^r u^t \\ \label{eq37}
&+& \frac{a^2}{\eta}\left[\left((u^t)^2- f\right)\partial_{t}\eta + u^r u^t \partial_{r}\eta\right]= 0
\end{eqnarray}
Once the expression for the speed of sound $a$ of the particular test fluid is known and after fixing the time derivatives to zero, the solution to the system of coupled equations~(\ref{eq36}) and (\ref{eq37}) provides us with the stationary fields $u^r(r)$, $\eta(r)$ and $a(r)$ and allow, in principle, the quantitative study of the spherical accretion onto a static spherically symmetric black hole. Assuming that the flow is smooth at all points of space-time, the stationary solutions of Eqs.~(\ref{eq36}) and (\ref{eq37}) are written as
\begin{equation}\label{eq38}
4\pi \bar{\mu} \eta u^r r^2= \dot{m},
\end{equation}
and
\begin{equation}\label{eq39}
\frac{d}{dr}\left[\ln \left(f u^t \right)\right]+\frac{a^2}{\eta}\frac{d \eta}{dr}=0,
\end{equation}
respectively, where the constant of motion $\dot{m}$ is to be identified as the matter flow rate.

The use of the perturbation scheme based on the continuity equation starts by considering linear perturbations to the stationary velocity and particle number density fields: $u^r(r,t) = u^r(r) + u^{r \prime}(r,t)$ and $\eta(r,t) = \eta(r) + \eta^\prime(r,t)$, where primed quantities stand for small time-dependent perturbations. We now define the variable $\psi=\eta u^r r^2$, whose stationary value $\psi(r)$ coincides, except for the constant $4 \pi \bar{\mu}$, with the steady mass accretion rate given by Eq.~(\ref{eq38}).

The first order perturbation to $\psi$ around stationary values is
\begin{equation}\label{eq40} 
\psi^\prime(r,t) = [u^r(r) \eta^\prime(r,t) + \eta(r) u^{r \prime}(r,t)] r^2,  
\end{equation}
and Eq.~(\ref{eq36}) acquires the form
\begin{equation}\label{eq41}
u^t \partial_{t} \eta^{\prime} 
+ \frac{\eta u^r}{f^2 u^t} \partial_{t} u^{r\prime}
= - \frac{1}{r^2} \partial_{r} \psi^{\prime}. 
\end{equation}
In general, the first order perturbation to the square of the speed of sound is given by
\begin{equation}\label{eq42}
a^{\prime 2} = a^2 + \frac{da^2}{d\eta}\eta^{\prime},
\end{equation}
The time evolution of $\eta^{\prime}$ and $u^{r\prime}$ follows from Eqs.~(\ref{eq40}) and (\ref{eq41}) and are given, respectively, by
\begin{equation}\label{eq43}
\partial_{t} \eta^{\prime} = - \frac{1}{r^2} 
\left(\frac{u^r}{f} \partial_{t} \psi^{\prime} 
+ f u^t \partial_{r} \psi^{\prime} \right),
\end{equation}
\begin{equation}\label{eq44}
\partial_{t} u^{r\prime} = \frac{f u^t}{\eta r^2}
\left(u^t \partial_{t} \psi^{\prime} 
+ u^r \partial_{r} \psi^{\prime} \right). 
\end{equation}
On the other hand, Eq.~(\ref{eq37}) is written in terms of perturbed quantities in the form
%
\begin{eqnarray}\nonumber
u^t \left[u^r \frac{a^2}{\eta}\partial_{t} u^{r\prime} 
+\partial_{t} \eta^{\prime} \right] + \partial_{r}\left(u^r u^{r\prime} \right) 
+ 2u^r u^{r\prime}\frac{a^2}{\eta} \partial_{r} \eta \\ \nonumber 
+(fu^t)^2 \partial_{r} 
\left(\frac{a^2}{\eta} \eta^{\prime} \right) = 0.  \\ \label{eq45}
\end{eqnarray}
Taking the time derivative of this last equation and substituting for the time derivatives of $\eta^{\prime}$ and $u^{r\prime}$ from Eqs.~(\ref{eq43}) and (\ref{eq44}), we obtain the differential equation obeyed by the perturbation to the mass accretion rate $\psi^\prime$
%
\begin{eqnarray}\nonumber
&\partial_{t}& \left(\eta h^{tt} \partial_{t} \psi^{\prime}
+ \eta h^{tr}\partial_{r} \psi^{\prime} \right) \\ \nonumber 
&+& \partial_{r} 
\left(\eta h^{rt}\partial_{t} \psi^{\prime}
+ \eta h^{rr}\partial_{r} \psi^{\prime} \right) \\ \nonumber
&=& (1 - 2a^2) \left(h^{rt}\partial_{t} \psi^{\prime}
+ h^{rr}\partial_{r} \psi^{\prime} \right)\frac{d\eta}{dr}, \\ \label{eq46}
\end{eqnarray}
where the coefficients $h^{\alpha\beta}$ are given by
\begin{eqnarray}\nonumber
h^{tt} &=& \frac{u^r (fu^t)}{f^2}
\left[(fu^t)^2 + u^r - (u^r)^2 a^2 \right], \\ \label{eq47} 
h^{tr} &=& h^{rt} = \frac{(u^r)^2(fu^t)^2}{f}(1 - a^2), \\
h^{rr} &=& u^r (fu^t) \left[(u^r)^2 - (fu^t)^2
a^2 \right]. \nonumber  
\end{eqnarray}

\subsection{\label{sec:sec5.1}Standing wave perturbation}
Concerning the study of perturbations in the form of a standing wave, it is has been already pointed out in the literature that, since an event horizon instead a physical surface occurs in a black hole, difficulties appear in fixing an appropriate inner boundary condition. Regularity of the flow at the black hole horizon singles out a unique solution, the Bondi one, which turns out to be transonic after the crossing of the sonic point. However, the standing wave perturbation should vanish even in the supersonic regime but there is no physical mechanism that allows to impose such a constraint. Since the standing wave analysis requires cancellation at the boundaries and also the continuity of the solution, we have to restrict ourselves to completely subsonic flows even though these may not entirely be representative of the precise manner of the infalling process. Consequently, in this section, we study the stability of a subsonic flow by assuming the trial standing wave perturbation 
\begin{equation}\label{eq48}
\psi^{\prime}(r,t) = \zeta(r) \exp\left(-{\rm i}\omega t\right),
\end{equation}
which, when substituted into Eq.~(\ref{eq46}), provides
\begin{widetext}
\begin{equation}\label{eq49}
\omega^2 h^{tt} \zeta^2 + {\rm i}\omega \left\{\frac{d}{dr} 
\left(h^{tr} \zeta^2 \right) - 2 h^{rt} \zeta^2 
\frac{d}{dr}\left[\ln (f u^t) \right]\right\}
+ \frac{h^{rr}}{(fu^t)^2} \frac{d\zeta}{dr} 
\frac{d}{dr}\left[\zeta (fu^t)^2\right] 
- \frac{d}{dr}\left(h^{rr}\zeta \frac{d \zeta}{dr}\right) = 0. 
\end{equation}
\end{widetext}
Integrating Eq.~(\ref{eq49}) over the radial coordinate with the integrated terms vanishing at the boundaries, a dispersion relation for $\omega$ is obtained
\begin{equation}\label{eq50}
A \omega^2 - 2 {\rm i} B \omega + C = 0
\end{equation}
where
\begin{eqnarray}\nonumber
A &=& \int h^{tt} \zeta^2 dr,\\ \label{eq51}  
B &=& \int h^{rt}\zeta^2 \frac{d}{dr}
\left[ \ln \left(fu^t \right) \right] dr,  \\
C &=& \int \frac{h^{rr}}{(fu^t)^2} \frac{d\zeta}{dr}
\frac{d}{dr}\left[\zeta (fu^t)^2 \right]dr. \nonumber  
\end{eqnarray}
The roots of Eq.~(\ref{eq50}) are given by
\begin{equation}\label{eq52}
\omega = {\rm i} \frac{B}{A} \pm {\rm i} \sqrt{\frac{B^2}{A^2}+\frac{C}{A}}.
\end{equation}
Clearly, the sign of the discriminant of the relation dispersion determines the stability of the standing wave. Now, from Eq.~(\ref{eq47}) it follows that $h^{rt}>0$. Also, taking into account that for infalling matter $(d\eta/dr)<0$ and referring to Eq.~(\ref{eq39}), we can conclude that $B>0$. Again, since $u^t=\sqrt{f+(u^r)^2}/f$ and for infalling fluid $u^r<0$, we have $h^{tt}<0$, which means $A<0$. Finally, it is easy verify that for subsonic flow $h^{rr}>0$. Consequently, $(B/A)<0$ and $(C/A)<0$, which implies whether an oscillatory and damped in time perturbation when $\vert C/A \vert > (B/A)$, or an overdamped perturbation if $\vert C/A \vert < (B/A)$. So, in any case, the stationary solution will be stable. 

Our goal, however, will be determine if quantum gravity effects enhance or diminish the dissipative effect associated to the coupling of the flow with the geometry of space-time. Notwithstanding, since the integrands in the coefficients in Eq.~(\ref{eq51}) depend in a complicated manner on the classical and quantum functions $f_0(r)$ and $f(r)$, the solution for $\omega$ is rather impractical for our purpose in the sense that it requires a considerable numerical effort after choosing a suitable mathematical distribution for modeling the amplitude of the standing wave. In the next section we will see that this aim can be achieved more easily by analyzing a disturbance in the form of traveling wave.

\subsection{\label{sec:sec5.2}Travelling wave perturbation}
In this case the perturbation will be modeled as a high-frequency travelling wave with a wavelength
much smaller than the horizon radius of the black hole \cite{6-PETT}. Then, the spatial part  $\zeta(r)$ of the perturbation $\psi^{\prime}(r,t)$ is written as a power series in $\omega$ in the form \cite{17-Naskar}
\begin{equation}\label{eq53}
\zeta_{\omega}(r) = {\rm exp} \left[\sum_{l=-1}^{\infty} \omega^{-l}k_l(r)\right].
\end{equation}
After replacing into Eq.~(\ref{eq49}), the coefficients of $\omega^2$ and $\omega$ can be collected and equated, each, to zero. In this way, two first order differential equations are obtained for $k_{-1}$ and $k_0$, respectively. Setting also to zero the coefficients of $\omega^0$, a second order differential equation for $k_1$, in terms of $k_{-1}$ and $k_0$, results. The solutions for $k_{-1}$ and $k_0$ reads
\begin{equation}\label{eq54}
k_{-1} = {\rm i} \int \left(h^{rr}\right)^{-1} 
\left[h^{tr}\pm\sqrt{\left(h^{tr}\right)^2-h^{rr}h^{tt}}\right]{\rm d}r,
\end{equation}
and
\begin{equation}\label{eq55}
k_0 = \ln \left\{(fu^t)^2\left[\sqrt{\left(h^{tr}\right)^2-h^{rr}h^{tt}}\right]^{-1}\right\}^{1/2},
\end{equation}
while the differential equation for $k_1$ is
\begin{widetext}
\begin{equation}\label{eq56}
2 \left(h^{rr} \frac{{\rm d}k_{-1}}{{\rm d}r} - {\rm i} h^{tr}\right)
\frac{{\rm d}k_1}{{\rm d}r} 
+ \frac{\rm d}{{\rm d}r} \left(h^{rr}\frac{{\rm d}k_0}{{\rm d}r}\right) 
+ h^{rr}\frac{{\rm d}k_0}{{\rm d}r} \frac{\rm d}{{\rm d}r} \left[k_0
- 2\ln(fu^t) \right] = 0 . 
\end{equation}
\end{widetext} 
That the power series can be truncated after these first three terms can be seen from their asymptotic behavior given by $k_{-1} \sim r$, $k_0 \sim \ln r$ and $k_1 \sim r^{-1}$, which also shows that the self-consistency requirement $\omega^{-l}\vert k_l(r) \vert \gg \omega^{-(l+1)}\vert k_{l+1}(r)\vert$ for the convergence of the series is fulfilled. So, we can write
\begin{equation}\label{eq57}
\zeta_{\omega}(r) \approx {\rm exp} \left[ \omega^1 k_{-1}(r)+k_0(r)+\omega^{-1}k_1(r) \right],
\end{equation}
It is then clear that $k_{-1}$ and $k_1$ only contribute to the phase of the travelling wave, and that the most important contribution to the amplitude of the perturbation $\psi^{\prime}(r,t)$ comes from $k_0$. A direct calculation gives
\begin{eqnarray}\nonumber
\vert\zeta_{\omega}(r)\vert &=& \chi \vert{\rm exp}\left[k_0(r) \right]\vert = \chi \left\lvert \frac{(fu^t)^2}
{\sqrt{\left(h^{tr}\right)^2-h^{rr}h^{tt}}}\right\rvert^{1/2} \\ \label{eq58}
&=& \chi \left[\frac{(u_t)^2}{(u^r)^2 a^2} \right]^{1/4}=\chi \left(\frac{1}{v^2 a^2} \right)^{1/4},
\end{eqnarray}
where $\chi$ is an arbitrary small real constant and we have used Eq.~(\ref{eq13}). It must be noted that, due to the fact that for spherically symmetric accretion $v$ never equals to zero and since $a$ is also different from zero, $\zeta_{\omega}$ never diverges so that the background solution is stable.

The dependence of $\vert\zeta_{\omega}(r)\vert$ on the metric coefficient $f$ makes evident the effect of the space-time geometry on the perturbation and determines the coupling between the accretion flow and the curvature of space-time. The differences between these effects in AS and in GR for the accretion of isothermal fluids are already evident from Eq.~(\ref{eq58}) because a higher value of the three-velocity of the accretion $v$ is always accompanied by a lower value of $\vert\zeta_{\omega}(r)\vert$ and vice versa. The variation of $\vert\zeta_{\omega}(r)\vert$ with the radial distance to the black hole can be calculated by writing the Hamiltonian for each type of isothermal fluid as a polynomial equation in the three-velocity $v$. After fixing an appropriate value for the Hamiltonian, this equation can be solved and the result can be inserted into Eq.~(\ref{eq58}). These polynomial equations for each value of the state parameter $k$, except for $k=1$ for which the expression for $v^2$ is immediate, read:

\begin{equation}\label{eq59}
w^2 - w + \frac{f(r)}{{\cal{H}}^2(r)r^4} = 0 \qquad (k=1/2),
\end{equation}
where $w=v^2$;

\begin{equation}\label{eq60}
v^3 - v + \frac{f(r)}{{\cal{H}}^{3/2}(r)r^2} = 0 \qquad (k=1/3),
\end{equation}
and
\begin{equation}\label{eq61}
x^4 - x - \frac{f(r)}{{\cal{H}}^{4/3}(r)r^{4/3}} = 0 \qquad (k=1/4),
\end{equation}
where $x=v^{2/3}$.

\begin{figure}[!ht]
	\centering
	\minipage{0.45\textwidth}
	\includegraphics[width=5.9cm,height=5.4cm]{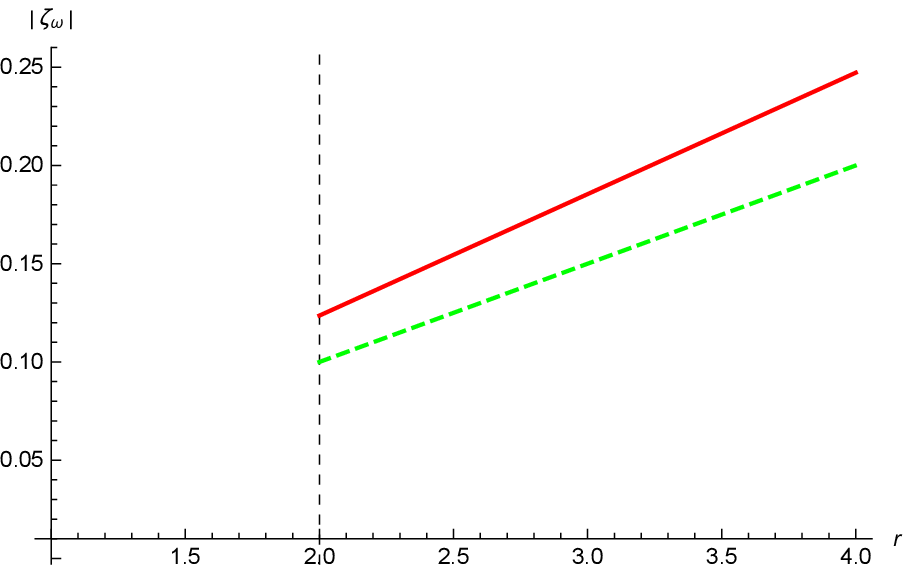}
	\label{F7_1}
	\endminipage\hfill
	\minipage{0.45\textwidth}
	\includegraphics[width=5.9cm,height=5.4cm]{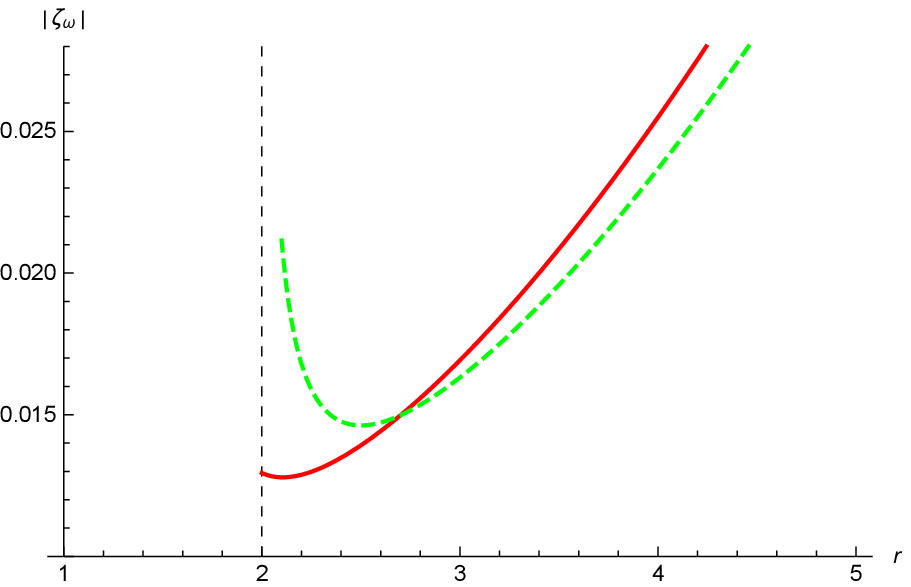}
	\label{F7_2}
	\endminipage\hfill\\
	\minipage{0.45\textwidth}
	\includegraphics[width=6.4cm,height=5.4cm]{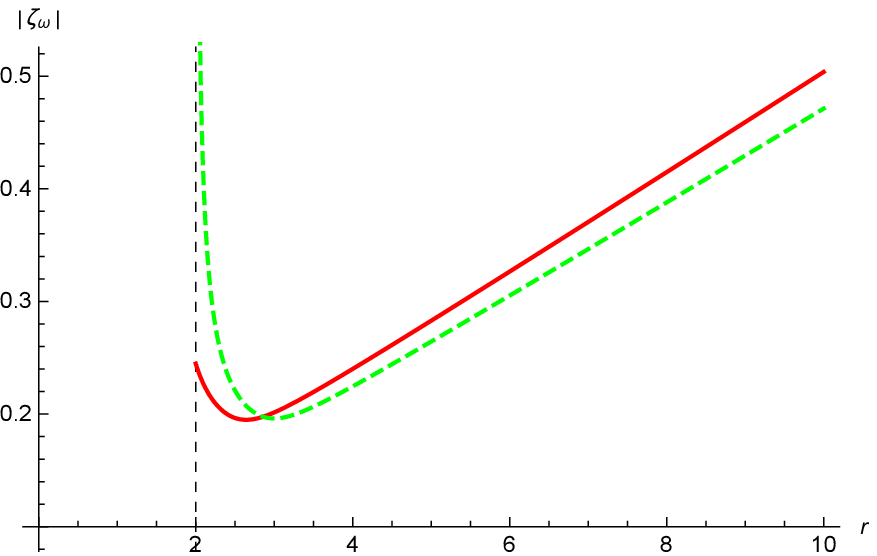}
	\label{F7_3}
	\endminipage\hfill
	\minipage{0.45\textwidth}
	\includegraphics[width=6.5cm,height=5.4cm]{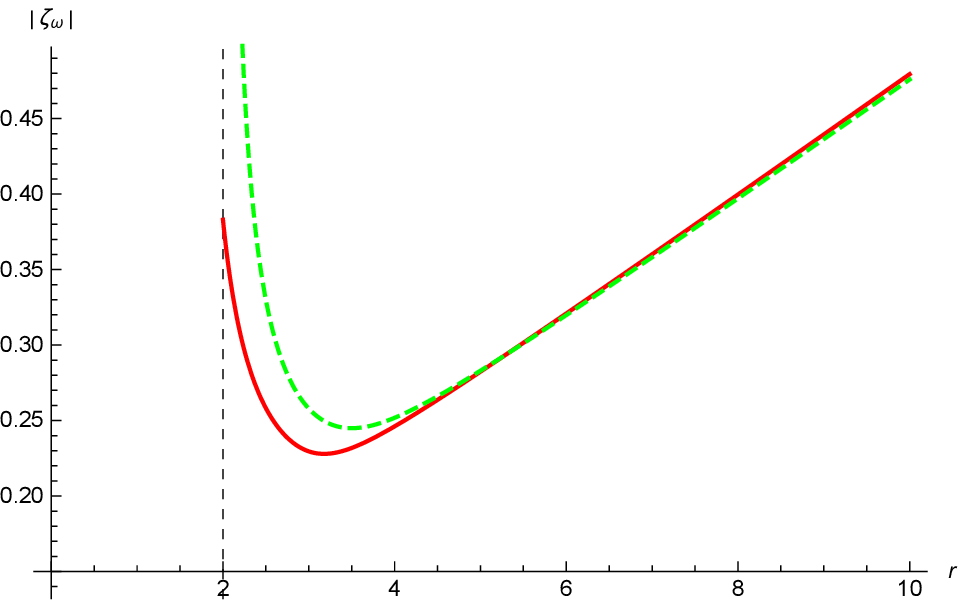}
	\label{F7_4}
	\endminipage\hfill
	\caption{Behavior of the amplitude of the travelling wave perturbation $\vert\zeta_{\omega}(r)\vert$ in AS (red) versus GR (green) for accretion of ultra-stiff fluid (upper left panel), ultra-relativistic fluid (upper right panel), radiation fluid (lower left panel), and sub-relativistic fluid (lower right panel). The vertical dashed line stands for the horizon radius $r_{hS}$ of the classical Schwarzschild black hole. The values of the parameters are: ${\cal{H}}^{AS}={\cal{H}}_c^{AS}+0.01$, ${\cal{H}}^{GR}={\cal{H}}_c^{GR}+0.01$, $M=1$, $\xi=0.5$, $\chi=0.1$.} \label{Fig7}
\end{figure}

From the solutions to the previous equations for each value of $k$, we have selected only the one corresponding to infalling or outflowing matter with velocity going to zero at the infinite spatial. Upon replacing the chosen solutions into Eq.~(\ref{eq58}) we can construct the plots shown in Fig.~\ref{Fig7} for subsonic flow (similar plots can also be done for transonic and supersonic flows). In this figure, the upper left panel shows that for ultra-stiff fluid the amplitude of the perturbation is greater in AS as compared with the amplitude in GR for all values of $r$ ($r>r_{hS}$) that is, for the same radial distance a locally static observer measures a greater amplitude in AS than in GR thus leading to a lower stability of the accretion process in AS. This means that, for ultra-stiff fluid, the coupling of the infalling matter with the space-time geometry is weaker in AS than in GR, a result which is undoubtedly associated to the anti-screening character of the gravitational interaction in AS. For the remaining cases of isothermal accretion namely $k=1/2$, $k=1/3$ and $k=1/4$, the upper right panel, the lower left panel and the lower right panel in Fig.~\ref{Fig7} show, respectively, that the amplitude of the perturbation in AS is detracted from its value in GR if $r_{hS}<r<r_{\mathrm{cross}}$ and it is instead enhanced if $r>r_{\mathrm{cross}}$, where the value of $r_{\mathrm{cross}}$ for each type of isothermal fluid is simply obtained by solving for $r$ the equation that equals the quantum amplitude with the classical one. This means that the coupling of the fluid with the space-time curvature, which acts on the perturbation in the manner of a dissipative effect, is stronger in AS than in GR for $r<r_{\mathrm{cross}}$ and is weaker for $r>r_{\mathrm{cross}}$. It is worthy to note that for sub-relativistic fluid the main effect is the reduction of the amplitude of the travelling wave perturbation with the consequent increase in the coupling of the fluid with the curvature of space-time, a result that has been already reported in \cite{31-SOSA}.
\section{\label{sec:sec7}Conclusions}
In this work we have reviewed the problem, previously analyzed in Ref.~\cite{27-FAYJ}, of the steady accretion onto a R-G improved Schwarzschild black hole in the AS theory with higher derivatives. In this sense the accretion process is considered as a quantum correction to the known spherically symmetric accretion in general relativity. As in \cite{27-FAYJ}, we have used an isothermal fluid as our test fluid and we have specialized to the cases: ultra-stiff fluid ($k=1$), ultra-relativistic fluid ($k=1/2$), radiation fluid ($k=1/3$), and sub-relativistic fluid ($k=1/4$). Since our interest has been to describe the most general aspects of the problem, we have neglected viscous effects, heat transport, fluid self-gravity, and effects associated to the back-reaction of the fluid on the geometry. By using the Hamiltonian dynamical system procedure introduced in \cite{29-CHA-SA}, we have recovered the results by the authors in \cite{27-FAYJ} for the accretion of ultra-stiff, radiation and sub-relativistic fluids but, at difference to what established in that work, we have found that accretion of ultra-relativistic fluids are a physical possibility in AS. In fact, for this type of fluid, as for the others considered here, we have found not only a shifting of the orbits of the Hamiltonian in the $(v,r)$ plane towards the central object and an enhancement of the flow velocity near the black hole as compared to what happens in GR, but also subsonic, supersonic and transonic regimes. This is an important result because there seems to be no physical reason for the no physical significance of the accretion of an ultra-relativistic fluid in AS. As expected, the quantum effects on the accretion are fully governed by the parameter $\xi$ in the RG-improved metric coefficient $f(r)$ in Eq.~(\ref{eq2}) with the classical behavior recovered in the limit $\xi \rightarrow 0$.

We have also analyzed the issue of the stability of the accretion and we have contrasted the results with the known stability of the accretion flow in classical general relativity. Using a perturbative procedure based on the continuity equation, the mass accretion rate of the infalling matter, considered as an isothermal fluid, has been subjected to small linear perturbations in the form of a standing wave and in the form of a travelling wave. For the standing wave perturbation we can conclude that, as in the general relativistic realm, the disturbance is damped in time so that the accretion is stable. As for the travelling wave perturbation a simple criterion, based on the behavior of the coefficient $k_0$ in the series expansion of the spatial part of the perturbation, has allowed us to compare between the quantum and the classical frameworks, and to reach to the conclusion that, where quantum gravity effects will have to be accounted for, the amplitude of the travelling wave perturbation for ultra-stiff fluid becomes enhanced as compared with the classical one. This indicates that the coupling of this type of fluid with the space-time curvature is weaker in AS than in GR. For ultra-relativistic, radiation and sub-relativistic fluids, the amplitude of the perturbation is reduced or enhanced depending on whether the local observer is located in the immediate neighborhood of the black hole horizon or not, respectively. The value of the radial coordinate $r_{\mathrm{cross}}$ at which this transition takes place can be easily calculated by solving for $r$ the equation that equals the classical amplitude with the quantum one. This means that the coupling of the accreting fluid with the geometry of the space-time, which is responsible for the damping of the perturbation, is stronger in the quantum case than in the classical one for $r<r_{\mathrm{cross}}$ and weaker for $r>r_{\mathrm{cross}}$. Since the disturbed quantity it has been the mass accretion rate, a physical quantity which is in principle measurable, it would be very interesting to study the possible observable effects that could be used to test asymptotic safety in the future. 

\section*{ACKNOWLEDGEMENTS}
We acknowledge financial support from COLCIENCIAS through the project with code 50754 associated to the CONVOCATORIA 757 PARA DOCTORADOS NACIONALES 2016. We also acknowledge financial support from Universidad Nacional de Colombia through the project with code 36031 within the program for the institutional strengthening of scientific research.


\begin{thebibliography}{}
\addcontentsline{toc}{chapter}{Bibliography}

\bibitem[1]{1}
http://www.eventhorizontelescope.org

\bibitem[2]{2}
http://www.mpe.mpg.de/ir/gravity

\bibitem[3]{3-Bondi}
H. Bondi, Mon. Not. R. Astron. Soc. \textbf{112}, 195 (1952).

\bibitem[4]{4-Michel}
F.C. Michel, Astrophys. Space Sci. \textbf{15}, 153 (1972).

\bibitem[5]{5-various}
M. Begelman, Astron. Astrophys. \textbf{70}, 583 (1978); 
L.I. Petrich, S.L. Shapiro, and S.A. Teukolsky, Phys. Rev. Lett. \textbf{60}, 1781 (1988);
J. Karkowski, B. Kinasiewicz, P. Mach, E. Malec, and Z. Swierczynski, Phys. Rev. D \textbf{73}, 021503 (2006); 
M. Jamil, M.A. Rashid, and A. Qadir, Eur. Phys. J. C \textbf{58}, 325 (2008);
S.B. Giddings and M.L. Mangano, Phys. Rev. D \textbf{78}, 035009 (2008); 
E. Babichev, S. Chernov, V. Dokuchaev, and Yu. Eroshenko, Phys. Rev. D \textbf{78}, 104027 (2008); J.A. Jim\'enez Madrid and P.F. Gonz\'alez-D\'\i az, Grav. \& Cosmol. \textbf{14}, 213 (2008);
M. Sharif and G. Abbas, Mod. Phys. Lett. A \textbf{26}, 1731 (2011);
V.I. Dokuchaev and Y.N. Eroshenko, Phys. Rev. D \textbf{84}, 124022 (2011); 
J. Bhadra and U. Debnath, Eur. Phys. J. C \textbf{72}, 1912 (2012); 
E. Babichev, V. Dokuchaev, and Yu. Eroshenko, Class. Quant. Grav. \textbf{29}, 115002 (2012); 
A.J. John, S.G. Ghosh, and S.D. Maharaj, Phys. Rev.  D \textbf{88}, 104005 (2013); 
P. Mach and E. Malec, Phys. Rev. D \textbf{88}, 084055 (2013);
A. Ganguly, S.G. Ghosh, and S.D. Maharaj, Phys.Rev. D \textbf{90}, 064037 (2014). 

\bibitem[6]{6-PETT}
J.A. Petterson, J. Silk, and J.P. Ostriker, Mon. Not. R. Astron. Soc. \textbf{191}, 571 (1980).

\bibitem[7]{7-TD}
T. Theuns and M. David, Astrophys. J. \textbf{384}, 587 (1992).

\bibitem[8]{8-Garlick}
A.R. Garlick, Astron. Astrophys. \textbf{73}, 171 (1979).

\bibitem[9]{9-Gaite}
J. Gaite, Astron. Astrophys. \textbf{449}, 861 (2006).

\bibitem[10]{10-Moncrief}
V. Moncrief, Astrophys. J. \textbf{235}, 1038 (1980).

\bibitem[11]{11-Malec}
E. Malec, Phys. Rev. D \textbf{60}, 104043 (1999).

\bibitem[12]{12-DS}
T.K. Das and A.Sarkar, Astron. Astrophys. \textbf{374}, 1150 (2001).

\bibitem[13]{13-MRD}
I. Mandal, A.K. Ray, and T.K. Das, Mon. Not. R. Astron. Soc. \textbf{378}, 1400 (2007).

\bibitem[14]{14-Ray}
A.K. Ray, Mon. Not. R. Astron. Soc. \textbf{344}, 1085 (2003).

\bibitem[15]{15-RR}
N. Roy and A.K. Ray, Mon. Not. R. Astron. Soc. \textbf{380}, 733 (2007).

\bibitem[16]{16-MM}
P. Mach and E. Malec, Phys. Rev. D \textbf{78}, 124016 (2008).

\bibitem[17]{17-Naskar}
T. Naskar, N. Chakravarty, J.K. Bhattacharjee, and A.K. Ray, Phys. Rev. D \textbf{76}, 123002 (2007);
D.B. Ananda, S. Bhattacharya and T.K. Das, GRG 47, 96 (2015).

\bibitem[18]{18-Weinberg}
S. Weinberg in {\it General Relativity: an Einstein Centenary Survey}, S.W. Hawking and
W. Israel (Eds.), Cambridge University Press (1979).

\bibitem[19]{19-Reuter}
M. Reuter, Phys. Rev. D \textbf{57}, 971, (1998).

\bibitem[20]{20-various2}
D. Dou and R. Percacci, Class. Quant. Grav. \textbf{15}, 3449 (1998);
W. Souma, Prog. Theor. Phys. \textbf{102}, 181 (1999);
O. Lauscher and M. Reuter, Class. Quant. Grav. \textbf{19}, 483 (2002); 
M. Reuter and F. Saueressig, Phys. Rev. D \textbf{65}, 065016 (2002);
A. Codello, R. Percacci and C. Rahmede, Int. J. Mod. Phys. A \textbf{23}, 143 (2008);
P.F. Machado and F. Saueressig, Phys. Rev. D \textbf{77}, 124045 (2008);
D. Benedetti, P.F. Machado and F. Saueressig, Mod. Phys. Lett. A \textbf{24}, 2233 (2009);
Nucl. Phys. B \textbf{824}, 168 (2010);
M.R. Niedermaier, Phys. Rev. Lett. \textbf{103}, 101303 (2009);
A. Eichhorn, H. Gies and M.M. Scherer, Phys. Rev. D \textbf{80}, 104003 (2009); 
K. Groh and F. Saueressig, J. Phys. A \textbf{43}, 365403 (2010);
A. Eichhorn and H. Gies, Phys. Rev. D \textbf{81}, 104010 (2010); 
S. Nagy, J. Krizsan and K. Sailer, JHEP \textbf{1207}, 102 (2012);
D. Benedetti and F. Caravelli, JHEP \textbf{1206}, 017 (2012) [Erratum-ibid. \textbf{1210}, 157 (2012)];
M. Demmel, F. Saueressig and O. Zanusso, JHEP \textbf{1211}, 131 (2012);
J.A. Dietz and T.R. Morris, JHEP \textbf{1301}, 108 (2013);
A. Eichhorn, Phys. Rev. D \textbf{87}, 124016 (2013).

\bibitem[21]{21-BoRe}
A. Bonanno, M. Reuter, Phys. Rev. D \textbf{62}, 043008 (2000).

\bibitem[22]{22-ReTu}
M. Reuter and E. Tuiran, Phys. Rev. D 83, 044041 (2011).

\bibitem[23]{23-Ko-Sa}
B. Koch and F. Saueressig, Class. Quan. Grav. \textbf{31}, No. 1, 015006 (2014).

\bibitem[24]{24-Go-Ko}
C. Gonz\'alez, B. Koch, Int. J. Mod. Phys. A \textbf{31}, No. 26, 1650141 (2016).

\bibitem[25]{25-various3}
B.F.L. Ward, Acta Phys. Polon. B \textbf{37}, 1967 (2006); 
K. Falls, D.F. Litim and A. Raghuraman, Int. J. Mod. Phys. A \textbf{27}, 1250019 (2012); 
S. Basu and D. Mattingly, Phys. Rev. D \textbf{82}, 124017 (2010); 
R. Casadio, S.D.H. Hsu and B. Mirza, Phys. Lett. B \textbf{695}, 317 (2011); 
K. Falls,  {\it Asymptotic Safety and Black Holes}, Springer theses, Springer (2013).

\bibitem[26]{26-Yang}
R-J. Yang, Phys. Rev. D \textbf{92}, 084011 (2015).

\bibitem[27]{27-FAYJ}
M.U. Farooq , A.K. Ahmed, R-J. Yang and M. Jamil, Chinese Physics C, \textbf{44}, 065102 (2020).

\bibitem[28]{28-Cai-Easson}
Y-F. Cai and D.A. Easson, JCAP \textbf{09}, 002 (2010).


\bibitem[30]{29-CHA-SA}
E.Chaverra and O. Sarbach, Class, Quant. Grav. \textbf{32}, 15 (2015);
E.Chaverra, M.D. Morales and O. Sarbach, Phys. Rev. D \textbf{91}, 104012 (2015);
E.Chaverra, P. Mach and O. Sarbach, Class, Quant. Grav. \textbf{33}, 10 (2016).

\bibitem[31]{30-AHMED}
A.K. Ahmed, M. Azreg-Ainou, M. Faizal and M. Jamil, Eur. Phys. J. C \textbf{76}, 280 (2016);
A.K. Ahmed, M. Azreg-Ainou, S.Bahamonde, S. Capozziello and M. Jamil, Eur. Phys. J. C \textbf{76}, 269 (2016).


\bibitem[32]{31-SOSA}
L.D. Sosapanta, {\it Efectos cu\'anticos gravitacionales sobre la estabilidad de la acreci\'on esf\'ericamente sim\'etrica hacia un agujero negro de Schwarzschild}, M.Sc. Thesis, Universidad Nacional de Colombia, Medell\'\i n Campus (2017), https://repositorio.unal.edu.co/bitstream/handle/unal/64741/1086361331.2018.pdf?sequence=1

\end{thebibliography}
\end{document}